\documentclass[12pt]{iopart}
\usepackage{graphicx}
\usepackage{ulem}
\usepackage{color}
\usepackage{iopams}
\usepackage{setstack}
\usepackage{mathabx}

\begin{document}

\title{Free energy and inference in living systems}
\author{Chang Sub Kim}
\address{Department of Physics,
Chonnam National University,
Gwangju 61186, Republic of Korea}
\ead{cskim@jnu.ac.kr}

\begin{abstract}
Organisms are nonequilibrium, stationary systems self-organized via spontaneous symmetry breaking and undergoing metabolic cycles with broken detailed balance in the environment.
The thermodynamic free-energy principle describes an organism's homeostasis as the regulation of biochemical work constrained by the physical free-energy cost.
In contrast, recent research in neuroscience and theoretical biology explains a higher organism's homeostasis and allostasis as Bayesian inference facilitated by the informational free energy.
As an integrated approach to living systems, this study presents a free-energy minimization theory overarching the essential features of both the thermodynamic and neuroscientific free-energy principles.
Our results reveal that the perception and action of animals result from active inference entailed by free-energy minimization in the brain, and the brain operates as Schr{\"o}dinger's machine conducting the neural mechanics of minimizing sensory uncertainty.
A parsimonious model suggests that the Bayesian brain develops the optimal trajectories in neural manifolds and induces a dynamic bifurcation between neural attractors in the process of active inference.

\end{abstract}

\vskip1in
\hskip1in\today
\maketitle

\section{Introduction}
\label{Introduction}
Although there is no standard definition of life \cite{Nurse2020,Ivanitskii2010,Goldenfeld2011,Friston2013a,Farnsworth2018,Fang2019,Kauffman2020}, literature often states that a living system tends to reduce its entropy, defying the second law of thermodynamics to sustain its nonequilibrium (NEQ) existence.
However, conforming to the second law of thermodynamics, adjudication between the entropy reduction and augmentation of an open system must depend on the direction of irreversible heat flux at the system-reservoir interface.
Organisms are open systems in the environment; hence, they obey the second law by contributing to the total-entropy increase in the universe.
The above confusion, perhaps, is rooted in Erwin Schr\"odinger's annotation, which metaphorically explains living organisms as feeding on negative entropy \cite{Schrodinger1944}.
In the same monograph, Schr\"odinger continues to explicate that a more appropriate discussion for metabolism is to be addressed in terms of free energy (FE).
He made this clarification because, in contrast to the Clausius entropy to which he was referring, thermodynamic FE always decreases during irreversible processes in any open system \cite{Landau1980}.
Many studies have been based on Schr\"odinger's insight into how biological systems can be explained by physical laws and principles.
We examine the definition of life in terms of FE minimization.

Organisms maintain biologically essential properties, such as body temperature, blood pressure, and glucose levels, which are distinct from ambient states.
Living systems continuously exchange heat and material fluxes with the environment by performing metabolic work, which is subject to the energy balance described by the first law of thermodynamics.
The second law posits that the entropy of an isolated macroscopic system increases monotonically with any spontaneous changes.
Organisms and the environment together constitute the biosphere, which is isolated and macroscopic; thus, metabolic processes in organisms increase the \textit{total} entropy.
The second law affects organisms by limiting metabolic efficiency.
The thermodynamic free-energy principle (TFEP) encompasses thermodynamic laws and provides qualitative and quantitative explanation of how living systems biophysically sustain homeostasis by minimizing FEs.
Recent studies have addressed the modern metabolism perspective as energy regulation of multisensory integration across both interoceptive and exteroceptive processes \cite{Corcoran2017,Quigley2021}.
This explains metabolism not only at the level of individual organisms, but also at the ecosystem and planetary levels \cite{Falkowski2016,Goldford2018}, and emphasizes the energetics and power efficiency in brain performance \cite{Sengupta2013,Balasubramanian2015,Levy2021}.
In contrast, the ability of organisms to undergo allostasis, which predictively regulates homeostasis \cite{Sterling2012,Schulkin2019}, or, more generally, their autopoietic properties \cite{Maturana1980}, are unable to be explained by the TFEP.
Allostatic ability is the main driver of adaptive fitness, the emergence of which cannot be solely attributed to a (bio)physical self-organization from a myriad of emergent possibilities in the primitive circuits of neuronal activities.
Organisms are under environmental constraints, and adaptive fitness, or natural selection, is the consequence of survival optimization in specific environments during evolution.
Therefore, the FE minimization scheme requires a top-down or high-level computational mechanism that facilitates hardwiring of the allostatic capability.

The brain-inspired free-energy principle (FEP) in neuroscience and theoretical biology suggests a universal biological principle in an axiomatic manner, which provides the informational FE-minimization formalism, that accounts for the perception, learning, and behavior of living systems \cite{Friston2009,Friston2010}.
This principle has been also applied to other cognitive systems, such as artificial intelligence and robots \cite{Catal2019,Matsumoto2020,Sancaktar2020,Catal2021,Meo2021,Mazzaglia2022,LancelotDaCosta2022}; however, our study primarily focuses on living systems and implications of the FEP in a biological context, emphasizing the embodied nature of inference \cite{Pezzulo2015}.
According to the informational free-energy principle (IFEP), all life forms are evolutionarily self-organized to minimize \textit{surprisal}, which is an information-theoretic measure of the improbability or unexpectedness of the environmental niche of organisms.
Informational FE (IFE) is a mathematical construct, rather than a physical (thermodynamic) quantity, specified by the temperature, chemical potential, volume, etc.
Informational FE mathematically bounds the surprisal from above; accordingly, the IFEP suggests that natural selection reflects minimization of IFE in an organism as a proxy for surprisal at all biological time scales.
The IFEP employs Helmholtz's early idea of perception as unconscious inference \cite{Helmholtz}:
an organism's brain possesses an internal model of sensory generation and infers the external causes of sensory data by matching them with prior knowledge.
The active-inference framework following from the IFEP encapsulates motor control and planning beyond Helmholtzian perception as an additional inferential scheme \cite{Friston2011,Friston2013b}.
The brain possesses the probabilistic internal model whose parameters (sufficient statistics) are encoded by brain variables in the NEQ stationary state; however, thus far, no physical theory has been developed for determining NEQ probabilities in the macroscopic brain.
In practice, the IFEP assumes open forms, or some fixed forms, for the NEQ densities and implements IFE minimization.
The Gaussian fixed-form assumption can be used to convert the IFE to a sum of discrepancies between the predicted and actual signals \cite{BuckleyKim2017}, which is known as \textit{prediction error} in predictive coding theory \cite{Rao2011}.
Commonly, the transformed IFE objective is minimized by employing the gradient-descent method widely used in machine learning \cite{Sutton1998}.
The resulting variational-filtering equations compute the Bayesian inversion of sensory data by inferring the external sources \cite{Balaji2011}, known as recognition dynamics (RD) \cite{Friston2009}.
Recently, the IFEP was generalized in a manner that minimizes \textit{sensory uncertainty}, which is a long-term surprisal over a temporal horizon of an organism's changing environmental niche \cite{Kim2018}.
Despite being a promising universal biological principle, the IFEP has led to controversy regarding its success as the universal principle and its distance between biophysical reality and epistemological grounds  \cite{Fiorillo2010,Kirchhoff2018,Colombo2018,Ramstead2018,Sanchez2021,Korbak2021,Biehl2021,Aguilera2021,Bruineberg2021,Raja2021}.

In this study, the two FE approaches are jointly considered to develop a unified paradigm for living systems: the TFEP does not describe the brain's ability to infer and act in the environment, whereas the brain-inspired IFEP is mainly a purposive (hypothesis-driven) framework lacking intimate connections to neuronal substrates and physical laws.
Our goal is to link the two FEPs and propose a biological FEP that integrates the reductionistic base and top-down teleology in the brain.
In addition, we unveil the attractor dynamics that computes allostatic regulation, perception and motor inference, in the brain, based on our proposed FE-minimization framework.
A similar approach was reported in \cite{Friston2019:particular}, in which formalisms underwriting stochastic thermodynamics and the IFEP were presented without addressing the direct link between the thermodynamic and informational FE.
In addition, a unified Bayesian and thermodynamic view attempted to explain the brain's learning and recognition as a neural engine and proposed the laws of neurodynamics \cite{Shimazaki2020}.
We also note another recent work that made the neural manifold models from a symmetry-breaking mechanism in brain-network synergetics, commensurate with the maximum information principle \cite{Jirsa2022}.

In brain architecture, enormous degrees of freedom of neuronal activities pose the classical negligence in a high-dimensional problem; thus, the underlying neural dynamics appears to be stochastic.
However, we argue that perception, learning, and motor-inference in the brain is low-dimensional at the functional level, obeying the law of large numbers; accordingly, RD becomes deterministic, involving a limited number of latent variables.
For instance, a few joint angles suffice for the brain to infer arm movement in motor control.
In contrast, the emergence of deterministic RD is more subtle in perception and learning, which demands a systematic coarse-graining of stochastic neuronal dynamics.
Our investigation facilitates the systematic derivation of Bayesian-brain RD in terms of a few effective variables, which we term \textit{Bayesian mechanics} (BM); BM regulates the homeostasis and allostasis (that is, adaptive fitness) of living systems, conforming with the proposed biological FEP.

The concept of coarse-graining, or effective description, is ubiquitous in computational neurosciences \cite{Hopfield1982,Arami1997,Robinson1997,Deco2008,Fung2010,Transtrum2015,Cook2022}.
Here, we review the recent research relevant to our work, which motivated the development of BM.
Many previous studies of recorded neurons showed that population dynamics is confined to a low-dimensional manifold in empirical neural space, where trajectories are neural representations of the population activity \cite{Cunningham2014}.
In mathematical terms, the neural modes were defined as eigen-fields that span the neural manifold.
The latent variables, or collective coordinates, were defined as projection of the population activity onto the neural modes \cite{GallegoSolla2017,GallegoSolla2020}.
Other theoretical models support the idea that long-term dynamics in recurrent neural networks gives rise to the attractor manifold \cite{Monasson2013}, which is a continuous set of fixed points occupying a limited region of neural space.
Consequently, the attractor dynamics and switching between different attractors were manifested \cite{Monasson2015}, indicating a contextual change in neuronal representations \cite{Wills2005,Jezek2011}.
Moreover, the manifold hypothesis is widely applied in machine learning to approximate high-dimensional data using a small number of parameters \cite{CostaHero2004}.
Experimental studies showed that a dynamical collapse occurs in the brain from incoherent baseline activity to low-dimensional coherent activity across neural nodes \cite{Singer2021a,Singer2019,Singer2021b}.
Synchronized patterns emerged when the featured inputs and prediction derived from prior or stored knowledge matched; in contrast, when there was a mismatch, the high-dimensional multi-unit activity increased.
This observation also provided empirical evidence that neural signals reduce prediction errors, thereby minimizing the IFE.

Based on the results described above, we suggest that the latent dynamics can be effectively described by a small number of coarse-grained variables in the reduced dimension.
In this study, we formulate the BM of inferential regulation of homeostasis in living systems in terms of a few latent variables.
The latent variables are determined as the brain activities and their conjugate momenta that represent the external, environmental and motor, states and online prediction errors, respectively.
The sensory error at the peripheral level acts as a time-dependent driving source in BM, providing the neural mechanism for sensory, as well as motor, inferences.
Our continuous-state formulation in continuous time may be useful for studying situated-action problems in which biological systems must make decisions even during ongoing sensorimotor activity \cite{Cos2021}.

The remainder of this paper is organized as follows.
In Section~\ref{NEQFT}, we describe the establishment of the TFEP from NEQ fluctuation theorems when applied to living systems.
Section~\ref{stochastic dynamics} explains how stochastic dynamics at the neuronal level can be modeled and how a statistical approach can be used to determine the NEQ densities of neural states in the physical brain.
In Section~\ref{informational FEP}, we present the proposed biological IFEP minimizing long-term surprisal and establish its continuous-state implementation that yields BM in the neural phase space.
Next, in Section~\ref{Numerics}, we numerically integrate BM and manifest the attractor dynamics that performs perception and motor inference in the brain.
Finally, we summarize important outcomes of our investigation and the conclusions in Section~\ref{Ending}.
In Appendix, we present the dual closed-loop circuitry of active inference resulting from our model.

\section{Nonequilibrium fluctuation theorems applied to organisms}
\label{NEQFT}
Fluctuation theorems (FTs) concisely describe stochastic NEQ processes in terms of mathematical equalities \cite{JarPRL1997,Crooks_PRE1999}.
Although FTs were initially established for small systems, where fluctuations are appreciable, they also apply to macroscopic deterministic dynamics \cite{Kim2015}.
Here, we present FTs in an appropriate context of biological problems and propose that the FTs suggest a living organism is an NEQ system that maintains the \textit{housekeeping temperature}, $T$, (average $36.5~^\degree C$ in humans) within its body and employs metabolism \textit{isothermally} to act against its environment.

To this end, among the various representations of FTs, we use the NEQ work relation \cite{JarPRL1997}:
\begin{equation}
\label{FT1}
\langle e^{-\beta(W-\Delta F)}\rangle = 1,
\end{equation}
where $\beta=k_BT$, with $k_B$ being the Boltzmann constant and $T$ being the temperature as described below.
The mathematical equality given in Eq.~(\ref{FT1}) is known as the Jarzynski relation \cite{Seifert:EurPhys2008}.
Here, $W$ is the amount of experimental work performed on a small system immersed in a thermal reservoir and $\Delta F$ is the induced change in the Helmholtz FE of the system.
Accordingly, $W-\Delta F$ is the excess energy associated with each irreversible work process in the system, which is unavailable for a useful conversion.
The bracket, $\langle\cdots\rangle$, indicates the average over many work strokes, that is, work distribution subject to a protocol.
The average must be considered because the experimental work performance on small systems fluctuates.

The Jarzynski relation can be converted to an expression for entropy as follows.
By applying $\langle e^{-\beta W}\rangle \ge e^{-\beta\langle W\rangle}$ to Eq.~(\ref{FT1}), which is known as the Jensen inequality \cite{Cover-Thomas1991}, we obtain the inequality $\Delta F \le \langle W\rangle$.
This inequality is an alternative expression that can be used to apply the second law to \textit{isothermal} irreversible processes of the system initially prepared in equilibrium with a reservoir \cite{Kim2015}.
Using the inequality, one can consider the change in the average total entropy: $\langle\Delta S_{tot}\rangle = \langle\Delta S_{sys}\rangle + \langle\Delta S_{R}\rangle$,
where $\Delta S_{sys}$ is the change in the system entropy and $\Delta S_R$ is the change in the reservoir entropy.
The average associated with $\Delta S_R$, which is reversible by definition, can be further manipulated to obtain
$\langle \Delta S_R\rangle = -\langle Q_{sys}\rangle/T = (\langle W\rangle-\Delta U)/T$,
where $Q_R=-Q_{sys}$ is used in the first step, and then the thermodynamic first law is applied for $\langle Q_{sys}\rangle$ ; $U$ is the internal energy of the system.
Therefore, $T\langle \Delta S_{tot}\rangle = \langle W\rangle - (\Delta U-T\langle \Delta S_{sys}\rangle) = \langle W\rangle - \Delta F$, which leads to the stochastic second law for the combined small system and reservoir:
\begin{equation}
\label{Second1}
\langle \Delta S_{tot}\rangle \ge 0.
\end{equation}
The possibility of tightening the preceding inequality has been investigated among researchers by revealing a nonzero, positive bound, leading to thermodynamic uncertainty relations \cite{Horowitz2019,Seifert2019}.
The unavailable energy associated with individual work processes amounts to the total entropy change, namely, $\beta(W-\Delta F)=k_B^{-1}\Delta S_{tot}$ under isothermal conditions.
By applying the final identity to Eq.~(\ref{FT1}), the Jarzynski equality is cast to the integral form of entropy fluctuation:
\begin{equation}
\label{FT2}
\langle e^{-k_B^{-1}\Delta S_{tot}}\rangle = 1.
\end{equation}

In the biological context, $W$ is the amount of environmental work involved in the metabolism of a living system, such as the biological reactions of oxygenic photosynthesis and aerobic respiration \cite{Albarran-Zavala2007,Swedan2020}.
The biological work is not controllable and thus, stochastic.
The FTs describe the imbalance between energy intake and expenditure in an organism while maintaining the \textit{housekeeping temperature}.
The Helmholtz FE increment in the living system over a metabolic work cycle is limited by the average environmental work done on the organism.
The resulting inequality from the Jarzynski relation can be written in the organism-centric form as
\begin{equation}
\label{Second2}
\langle {\cal W}\rangle \le \Delta {\cal F},
\end{equation}
where we set $\langle{\cal W}\rangle=- \langle W\rangle$ and $\Delta{\cal F}\equiv -\Delta F$,
which now states that the work performance, $\langle{\cal W}\rangle$, of a biological system against the environment (e.g., via metabolism) is bounded from above by the thermodynamic FE cost, $\Delta{\cal F}$.
The preceding inequality reflects the limited efficiency of metabolic work in living systems.
Rare individual processes that violate Eq.~(\ref{Second2}) may occur in small systems; however, such statistical deflection is not expected in a finite biological system with macroscopic degrees of freedom.
The equality in Eq.~(\ref{Second2}) holds for reversible work cycles in inanimate matter, attaining thermodynamic efficiency at its maximum, but not in the metabolic processes of living organisms, which are irreversible.
Our consideration of metabolic work may be generalized to the multi-level autocatalytic cycles suggested as the chemical origins of life \cite{Lehman2021}.

Note here that we considered the temperature appearing in the Jarzynski relation as the body temperature of a specific biological system, unlike the usual implication of FTs;
in the standard derivation of the Jarzynski relation \cite{Jarzynski_AnnuRev2011}, the temperature, $T$, appearing in the NEQ equality is, by construction, the reservoir temperature.
The FT is generally intended for an irreversible process during which the system temperature may not be defined.
However, the initial and end states must be in equilibrium so that the FE is meaningful.
The subtlety lies in the fact that the end-state temperature may or may not be the same as the reservoir temperature for experiments performed in isolation after the initial equilibrium preparation.
Living organisms are in an NEQ stationary state, maintaining a housekeeping temperature, $T$, that is distinct from the ambient temperature, to which they equilibrate only when ceasing to exist.
Thus, organisms are viewed as isothermal systems, which are \textit{open} to heat and particle exchange with the environment.

The NEQ work relation expresses the second law of thermodynamics as the mathematical equality in Eq.~(\ref{FT1}).
The second law, in its biological context, renders the thermodynamic constraint on living organisms given by the inequality in Eq.~(\ref{Second2}), which reveals the inevitable (thermodynamic) FE waste produced during metabolic cycles.
However, this inequality accounts for neither self-adaptiveness nor brain functions, such as perception, learning, and behavior.
To address these essential features of life, researchers currently employ a hybridizing scheme, which first proposes how the system-level biological functions operate and then attempts to make connections to biophysical substances.
Particularly, the Bayesian mechanism built into the IFEP provides a crucial component in this promising hybrid explanation of life, which is described in detail in Section~\ref{informational FEP}.

\section{Statistical-physical description of the nonequilibrium brain}
\label{stochastic dynamics}
The brain is comprised of a myriad of complex neurons; accordingly, its internal dynamics at the mesoscopic level must obey some stochastic equations of motion on account of classical indeterminacy.
The relevant coarse-grained neural variables are local-scale population activities, or intra-area brain rhythms.
In the following, we consider that the brain matter itself constitutes the thermal environment at body temperature for the mesoscopic neural dynamics.

Below, we assume that the neural activity, $\mu$, at the coarse-grained population level obeys the stochastic dynamics \cite{KuboII}:
\begin{equation}
\label{Langevin1}
\frac{d\mu}{dt}=f(\mu;t) + w(t),
\end{equation}
where the inertial term in the Langevin equation was dropped by taking the over-damping limit.
Here, $f$ may represent both conservative and time-dependent metabolic forces, and $w$ represents random fluctuation characterized as a delta-correlated Gaussian noise satisfying the following conditions:
\[
\langle w(t)\rangle =0\quad {\rm and}\quad
\langle w(t)w(t^\prime)\rangle = I\delta(t-t^\prime),
\]
where $I$ is the noise strength.
In one dimension (1D), for simplicity, the environmental perturbation and noise strength are physically specified, respectively, as \cite{Risken}
\[ f=\frac{1}{m\gamma}A\quad{\rm and}\quad I=2\frac{k_BT}{m\gamma}, \]
where $A$ is a conservative force acting on a neural unit with mass, $m$, neglecting time-dependent driving, $T$ is the body temperature, and $\gamma$ is the phenomenological frictional coefficient whose inverse corresponds to momentum relaxation time.
The solutions to Eq.~(\ref{Langevin1}) describe the individual trajectories of random dynamical processes.

In general, colored noises can be considered beyond the delta-correlated white noise by generalizing Eq.~(\ref{Langevin1}) to incorporate the non-Markovian memory effect:
\[
m\int_{-\infty}^t dt^\prime\gamma(t-t^\prime)\dot\mu(t^\prime) = A(\mu) + \zeta.
\]
To ensure equilibrium at temperature $T$, the colored Langevin equation must satisfy the fluctuation-dissipation theorem that accounts for the nonsingular noise correlation \cite{Zwanzig}:
\[
\langle \zeta(t)\zeta(t^\prime)\rangle = 2k_BT\gamma(|t-t^\prime|).
\]
A standard example of such colored noise is the Orstein-Uhlenbeck memory kernel given by
$\gamma(|t-t^\prime|)= \gamma\tau^{-1} \exp(-|t-t^\prime|/\tau)$, where $\tau$ is the noise autocorrelation time.

As an alternative to the Langevin equation [Eq.~(\ref{Langevin1})], one may collectively consider an ensemble of identical systems displaying various values of the state, $\mu$, and ask how the statistical distribution changes over time.
After normalization, the ensemble distribution is reduced to the probability density, say $p(\mu,t)$, so that $p(\mu,t)d\mu$ specifies the probability that an individual Brownian particle is found in the range $(\mu, \mu+d\mu)$ at time $t$.
In the Markovian approximation, the change in the probability density is determined by the probability density at the current time, which is generally described by the master equation given in the continuous-state formulation as
\begin{equation}
\frac{\partial p(\mu,t)}{\partial t} = \int \Big\{ w(\mu,\mu^\prime)p(\mu^\prime,t)-w(\mu^\prime,\mu)p(\mu,t)\Big\}d\mu^\prime,
\end{equation}
where $w(\mu^\prime,\mu)$ is the transition rate of the state change from $\mu$ to another $\mu^\prime$.
We further assume that the transition occurs between two infinitesimally close states, $\mu$ and $\mu^\prime$, where $\mu^\prime-\mu = x \ll 1$, so that the transition rate sharply peaks at around $x=0$ to approximate the value as $w(\mu^\prime,\mu) \approx w(\mu;x)$.
Then, $p(\mu^\prime)$ can be expanded about $\mu$ to the second-order in $x$ and all higher-order terms are neglected.
Consequently, the master equation can be converted into the Smoluchowski-Fokker-Planck (S-F-P) equation \cite{Risken}:
\begin{equation}
\label{FP1}
\frac{\partial p(\mu,t)}{\partial t} = \frac{\partial}{\partial\mu}\Big\{ -D_1(\mu) +  \frac{\partial}{\partial \mu}D_2(\mu) \Big\} p(\mu,t),
\end{equation}
where, $D_1$ and $D_2$ correspond to the first two expansion coefficients in the Kramers-Moyal formalism, which are determined in the present case to be
\[ D_1 =f\quad{\rm and}\quad D_2 =\frac{1}{2}I.\]
The S-F-P equation can be expressed in three dimensions (3D) as
\begin{equation}
\label{FP2}
\frac{\partial p(\vec\mu,t)}{\partial t} + \nabla\cdot\Big\{ {\vec f}(\vec\mu) - D\nabla \Big\} p(\vec\mu,t)=0,
\end{equation}
where $\nabla$ is the gradient operation with respect to the three-dimensional state, $\vec\mu$.
In Eq.~(\ref{FP2}), the drift term, $p\vec f$, accounts for conservative potential forces.
In addition, the diffusion term, $D_2$, is denoted as $D$, assuming spatial isotropy, for simplicity and notational convention.

The S-F-P equation describes local conservation of the probability, $p(\vec\mu,t)$, in the state space spanned by the state vector, $\vec\mu$, which carries the probability flux, $\vec j$, identified as
\[ \vec j(\vec\mu,t) = p(\vec\mu,t){\vec f}(\vec\mu) - D\nabla p(\vec\mu,t). \]
In steady state (SS), $\partial p_{st}/\partial t=0$, where $p_{st}\equiv p(\mu,\infty)$; accordingly, the divergence of the SS flux, $\vec j_{st}\equiv\vec j(\mu,\infty)$, must vanish in the S-F-P equation:
\begin{equation}
\label{stdiv}
\nabla\cdot\vec j_{st}=0.
\end{equation}
If Brownian particles undergo motion in an isolated or infinite medium, $\vec j_{st}$ should disappear on the local boundary because the total flux through the surface must vanish to ensure probability conservation.\footnote{In an isolated or infinite medium, the net flux through the entire surface must vanish to ensure probability conservation, i.e., $\vec j_{net}\equiv\oint \vec j_{st}\cdot d{\vec a}=0$, where $d\vec a$ is the outward, infinitesimal area element. Accordingly, $\vec j_{st}=0$ at every point on the surface.}
Because the flux must be continuous over the entire space, the SS condition in Eq.~(\ref{stdiv}) imposes $\vec j_{st}\equiv 0$ everywhere, reflecting the \textit{detailed balance} between the drift flux and dissipative flux.
In this case, the system holds in equilibrium, where life ceases to exist.
The equilibrium probability can be obtained from the condition $j_{st}=0$, giving canonical Boltzmann probability as the result:
\[
p_{eq}(\mu) \propto \exp\{-\beta V(\mu)\},
\]
where $\beta=1/k_BT$ and $V(\mu)$ is potential energy.
The kinetic-energy term does not appear in $p_{eq}$ because the Langevin dynamics we consider are in the over-damping limit.

However, for a finite open system, such as a living organism, the system's SS flux does not necessarily vanish on the local boundary; instead, it must be compensated by the environmental afferent or efferent fluxes to achieve steady state.
Thus, for a living system, the detailed balance is not satisfied in the steady state \cite{Gnesotto2018,Lynn2021}; that is, $\vec j_{st}\neq 0$.
Instead, the vanishing condition of the divergence of the probability flux entails a necessary balance.
The mathematical expression in Eq.~(\ref{stdiv}) admits a non-vanishing vector field $\vec B(\vec\mu)$ via
\begin{equation}\label{flux2}
\vec j_{st}(\vec\mu)\equiv \nabla\times \vec B(\vec\mu),
\end{equation}
which shows that the SS flux is divergenceless or, equivalently, solenoidal \cite{Wang2008,Qian2013,Perl2021PRE}.
The life flux, $\vec j_{st}$, defined in this manner is unchanged when $\vec B$ is transformed to $\vec B^\prime = \vec B+\nabla \Lambda$, where $\Lambda$ is a scalar function of the state, $\vec \mu$.\footnote{The freedom to choose the vector field, $\vec B$, without affecting the physical quantity, $\vec j_{st}$, is known as \textit{gauge symmetry}. Recently, researchers attempted to determine the implication and utility of the gauge transformation in neuronal dynamics in the brain and emergent functions \cite{Sengupta2016,Sakthivadivel2022}.}
From Eq.~(\ref{flux2}), the following generalized balance condition must hold locally on the boundary:
\begin{equation}
\label{MBC}
p_{st}(\vec\mu){\vec f_{st}}(\vec\mu) = D\nabla p_{st}(\vec\mu) + \nabla\times \vec B(\vec\mu).
\end{equation}
The above \textit{modified detailed-balance} condition supports the frequent interpretation of the force field, $\vec f_{st}$, as the \textit{gradient flow} of the SS probability, $p_{st}$ \cite{Friston2013a,Parr2019}:
\begin{equation}\label{SSD}
{\vec f_{st}}(\vec\mu) = (D-Q)\nabla \ln p_{st}(\vec\mu),
\end{equation}
where we introduced the isotropic coefficient, $Q$, via
\[
Q\nabla\ln p_{st} \equiv - p_{st}^{-1}\nabla\times \vec B;
\]
for simplicity, the coefficient $Q$ is assumed to be isotropic as it was for the diffusion constant, $D$.
The gradient flow is driven by entropy because the most likely equilibrium state of the combined system and environment is achieved by maximizing the total entropy; hence, it is an entropic force, conforming to the second law.
Note that Eq.~(\ref{flux2}) mimics the Ampere law in magnetism \cite{Griffiths2017}; the effective field $\vec B$ may be construed as an induced field by the static current, $\vec j_{st}$.
Accordingly, the vector field, $\vec B(\vec\mu)$, can be determined by means of
\begin{equation}
\label{Biot-Savart}
\vec B(\vec\mu)=\frac{1}{4\pi}\int \vec j_{st}(\vec\mu^\prime)\times \frac{(\vec\mu-\vec\mu^\prime)}{|\vec\mu-\vec\mu^\prime|^3} d\vec\mu^\prime.
\end{equation}
Note that the modified detailed-balance condition given in Eq.~(\ref{MBC}) is only a formal description for determining the NEQ density, $p_{st}$, given SS flux, $\vec j_{st}$, or, equivalently, the environmental magnetic field, $\vec B$, in Eq.~(\ref{Biot-Savart}).
Precise determination of $p_{st}$ is an independent research subject, which may be non-Gaussian with a colored autocorrelation.

In general, it is difficult to obtain an analytic expression for the NEQ probability density for open systems, except in low-density and/or linear-response regimes \cite{Kim2009,Freitas2021}.
Because of morphological complexity, it is practically intractable to derive the NEQ densities specifying the physical brain states.
Accordingly, the neural states under continual sensory perturbation are assumed to be statistically described by time-dependent Gaussian densities, predicted from Gaussian random noises imposed on the Langevin description.

\section{Latent dynamics of sensorimotor inference in the brain}
\label{informational FEP}
Here, we present the BM for conducting Bayesian inversion of sensory observation in the brain under the proposed generalized IFEP.
This idea was previously developed by considering passive perception \cite{Kim2018} and only implicitly including active inference \cite{Kim2021}.
Here, we advance this formalism by explicitly introducing motor inference and planning in the generative models to fully conform to the active-inference framework.

The environmental states, $\vartheta$, generate sensory stimuli, $\varphi$, at the organism's receptors through mechanical, optical, or chemical perturbations, which are transduced in the brain's functional hierarchy in the form of a nervous signal.
The sensory perturbations may be altered by the organism's motor manipulation, and we designate $u$ to denote the motor variables responsible for such control over the effectors.
A crucial point here is that the brain has access only to the sensory data and not their causes; accordingly, from the brain's perspective, both the environmental states, $\vartheta$, and motor variables, $u$, are external, that is, \textit{hidden}.
In terms of these relevant variables, we define the variational IFE functional, denoted as $\cal F$:
\begin{equation}
\label{IFE}
{\cal F}[q(\vartheta,u),p(\varphi;\vartheta,u)] \equiv \int d\vartheta \int du~ q(\vartheta,u)\ln\frac{q(\vartheta,u)}{p(\varphi;\vartheta,u)},
\end{equation}
where $q(\vartheta,u)$ and $p(\varphi;\vartheta,u)$ are the recognition density (R-density) and generative density (G-density), respectively.
The R-density is the brain's online estimate of posterior beliefs about the external causes of the sensory perturbation (it probabilistically represents the environmental states).
The G-density encapsulates the brain's likelihood in beliefs about sensory-data generation and prior beliefs about the hidden environmental as well as motor dynamics (it probabilistically specifies the internal model of sensory-data generation, environmental dynamics, and motor feedback).
Note that whereas the R-density is the current estimate, the G-density contains the stored information in the brain, which can be updated by learning.
In this study, we generalize the R-density as a bi-modal probability of $\vartheta$ and $u$, and G-density as a tri-modal probability of $\vartheta$, $u$, and $\varphi$.
Note that a semicolon is used between the sensory perturbation, $\varphi$, and hidden variables $\vartheta$ and $u$ in the G-density rather than a comma to emphasize their differential role in perception.
The explicit inclusion of the motor variable, $u$, in the $q$ and $p$ densities is a key advancement over the standard definition of IFE \cite{BuckleyKim2017}.

Now, using the product rule, $p(\varphi;\vartheta,u)=p(\vartheta,u|\varphi)p(\varphi)$ for the G-density in Eq.~(\ref{IFE}), we decompose the IFE to a form applicable in the biological context:
\[
{\cal F}[q(\vartheta,u),p(\varphi;\vartheta,u)] = D_{KL}\left(q(\vartheta,u)\|p(\vartheta,u|\varphi)\right) - \ln p(\varphi),
\]
where $D_{KL}$ is the Kullback-Leibler divergence \cite{Cover-Thomas1991}.
Because $D_{KL}$ is non-negative, the following inequality holds, which underpins the IFEP described in Section~\ref{Introduction}:
\begin{equation}
-\ln p(\varphi) \le {\cal F}[q(\vartheta,u),p(\varphi;\vartheta,u)],
\end{equation}
where $-\ln p(\varphi)$ is the information-theoretic \textit{surprisal}.
Here, it is important to notice the resemblance between the preceding inequality and that given in Eq.~(\ref{Second2}) from the TFEP.

Under the IFEP, the organism's cognitive goal is to infer the hidden environmental causes of sensory inputs with feedback from the motor-behavior inference.
This goal is achieved by minimizing ${\cal F}$ with respect to the R-density, $q(\vartheta,u)$, which corresponds to the online adaptation of the sensory and motor modules in the brain.
For instance, in the classic reflex arc, the proprioceptive stimulus evokes the activity of sensory neurons in the dorsal root, and the motor variable is engaged by the effector's active states of the motor neurons in the ventral root.
The double procedures are involved in the minimization scheme to cope with the bi-modal cognitive nature of sensory and motor inferences: 1) the internal model is updated to better predict the sensory perturbation, and 2) the sensory perturbation is modified by the agent's motor engagement to further reduce the residual discrepancy with the internal model.
The former is termed as \textit{passive perception} and the latter as \textit{active perception}.
However, the two inferential mechanisms do not separately engage, but act as a whole in the sensorimotor closed loop in the embodied agent, and are therefore jointly termed as \textit{active inference} under the IFEP \cite{Friston2011,Friston2013b}.

To draw a connection between the IFE minimization and neuronal correlates, it is practically convenient to use the fixed form for the unknown R-density \cite{BuckleyKim2017}, whose sufficient statistics are assumed to be encoded neurophysiologically by brain variables, that is, neuronal activities.
Here, we write the R-density as $q(\vartheta,u)=q(\vartheta)q(u)$ by considering the external variables $\vartheta$ and $u$ as conditionally independent.
Furthermore, it is assumed that the factorizing densities, $q(\vartheta)$ and $q(u)$, are Gaussian; the means of the environmental states, $\vartheta$, and motor states, $u$, are encoded by the neuronal variables $\mu$ and $a$, respectively.
Then, by performing technical approximations similar to those used in \cite{BuckleyKim2017}, we convert the IFE \textit{functional}, $\cal F$, of the R- and G- densities to the IFE \textit{function}, $F$, of the neural representations $\mu$ and $a$, given sensory data, $s$.
The sensory data or inputs are a neural representation of the evoked perturbation, $\varphi$, at the receptors, detected by the organism's brain.
Here, the homunculus hypothesis, the brain as a neural observer, is implicit, which assumes teleological homology between the environmental processes and brain's internal dynamics.

The result for the IFE function, up to an additive constant, is given as
\begin{equation}
\label{Laplace-encoded1}
F(\mu,a;s)=-\ln p(s;\mu,a);
\end{equation}
here, the dependence on the second-order sufficient statistics, namely (co)variances of the R-density, was optimally removed.
Consequently, the brain must only update the means in the R-density in conducting the latent RD.
The mathematical procedure involved in Eq.~(\ref{Laplace-encoded1}) extends the Laplace approximation delineated in the review \cite{BuckleyKim2017}.
To complete the Laplace-encoded IFE, one must specify the inferential structure in the encoded G-density, $p(s;\mu,a)$.
We facilitate probabilistic implementation of the generative model using the product rule:
\begin{equation}
\label{Lap-G-density1}
p(s;\mu,a) = p(s|\mu,a)p(\mu,a),
\end{equation}
where the likelihood density, $p(s|\mu,a)$, is the brain's concurrent estimation of the encoded sensory data, $s$, from the neuronal response, $\mu$, and motor manipulation, $a$.
Assuming conditional independence between $\mu$ and $a$, the joint prior $p(\mu,a)$ can be further factorized as
\[p(\mu,a)=p(\mu)p(a),\]
where $p(\mu)$ and $p(a)$ are the brain's prior beliefs regarding the environmental-state changes and motor dynamics, respectively.
Thus, the Laplace-encoded IFE has been specified solely in terms of the neural variables $\mu$ and $a$, which is suitable for biologically-plausible implementation of active inference in the physical brain.

Sensory states, $s$, evoked by exogenous stimuli, neurophysically activate the neuronal population in the brain.
The population dynamics is complex and high-dimensional; however, the RD of the perceptual and behavioral inferences may be well-described in lower-dimensional neural manifolds.
Below, we formulate the generative equations of latent neural modes considering classical indeterminacy.
First, we assume that sensory data, $s$, encoded at the receptors are measured by a neural observer according to instant mapping:
\begin{equation}
\label{observation-eq}
s=g(\mu,a;\theta_g)+z,
\end{equation}
where $g$ is the generative model of the sensory data and $z$ is the observation noise.
The generative map, $g(\mu,a)$, encapsulates both the perceptual states, $\mu$, and motor states, $a$, which conjointly predict the sensory data, $s$.
We consider the sensory generative model as a continuous process of sensory prediction by $\mu$ and error prediction by $a$ via the effector alteration:
\[[s-g_1(\mu)]-g_2(a) \equiv s-g(\mu,a),
\]
where we set $g(\mu,a)=g_1(\mu)+g_2(a)$.
Second, we assume that the neural activity, $\mu$, obeys internal dynamics as described in Section~\ref{stochastic dynamics}:
\begin{equation}
\label{state-eq}
\frac{d\mu}{dt}=f(\mu;\theta_f)+w,
\end{equation}
where $f$ is the generative model of the neuronal change, and $w$ is the involved random noise.
Third, we assume that the motor state, $a$, bears the motor-neural dynamics:
\begin{equation}
\label{motor-eq}
\frac{da}{dt}=\pi(a;\theta_\pi)+\eta,
\end{equation}
where $\pi$ is the generative model of the motor-neuronal change and $\eta$ is the noise in the process.
The generative function, $\pi$, in Eq.~(\ref{motor-eq}) functions as the \textit{policy} in machine learning \cite{Sutton1998}:
the policy, $\pi(a;\theta_\pi)$, encapsulates the internal model of \textit{motor planning} in continuous time.
The dependence of the generative models on the parameters $\theta_g,\ \theta_f,{\rm and}\ \theta_\pi$ enables incorporation of a longer-term neural efficacy, such as synaptic plasticity; below, we omit the parameter dependence for notational simplicity.
For the neuronal generative equations, the continuous Hodgkin-Huxley model \cite{Kim2018} or a more biophysically realistic model can be employed; however, our simple model in Section~\ref{Numerics} suffices to unveil the emergence of BM.

Noises in the neural generative models [Eqs.~(\ref{observation-eq})-(\ref{motor-eq})] indicate stochastic mismatches between the cognitive objectives on the left-hand side (LHS) and their prediction through the generative functions/map.
Accordingly, we consider that $z$, $w$, and $\eta$ neurophysically encode the probabilistic generative models $p(s|\mu,a)$, $p(\mu)$, and $p(a)$, respectively, [Eq.~(\ref{Lap-G-density1})] in the neuronal dynamics.
Furthermore, we assume that the random noises are continuously distributed according to the normalized NEQ Gaussian.
Therefore, the Laplace-encoded likelihood, $p(s|\mu,a)$, and prior densities, $p(\mu)$ and $p(a)$, in Eq.~(\ref{Lap-G-density1}) can be assumed to take the following forms:
\begin{eqnarray}
\label{NEQ-densities}
&&p(s|\mu,a) = {\cal N}(s-g;0,\sigma_z),\nonumber\\
&&p(\mu) = {\cal N}(\dot\mu-f;0,\sigma_w),\\
&&p(a) = {\cal N}(\dot a-\pi;0,\sigma_\eta);\nonumber
\end{eqnarray}
here, ${\cal N}(x-h;0,\sigma)\equiv \exp\{-\frac{1}{2\sigma}(x-h)^2\}/{\sqrt{2\pi\sigma}}$ denotes a Gaussian density of stochastic variable $x-h$ with variance $\sigma$ about the zero mean\footnote{Here, we use $\sigma$, not $\sigma^2$, to denote the variances only to be consistent with the notations in an earlier publication \cite{BuckleyKim2017}.},
and $\dot x$ denotes the time derivative of $x$, that is, $dx/dt$.
The generative likelihood and prior densities in Eq.~(\ref{NEQ-densities}) are thought to be stationary solutions to the S-F-P equation or a more general non-Markovian extension, the biophysical derivation of which is beyond the scope of this work.
Instead, we assume the time-dependent Gaussian probabilities effectively at zero temperature as physically admissible densities encoding internal models in the brain.
Removing the assumption by deriving physical probability densities is a key theoretical demand in future studies.

Next, by substituting the expressions in Eq.~(\ref{NEQ-densities}) into Eq.~(\ref{Laplace-encoded1}) using the decompositions in Eq.~(\ref{Lap-G-density1}), we obtain an explicit expression for the IFE function at an instant $t$:
\begin{equation}
\label{Laplace-encoded2}
F(\mu,a;s) = \frac{1}{2\sigma_z}(s-g(\mu,a))^2 + \frac{1}{2\sigma_w}(\dot\mu-f(\mu))^2 + \frac{1}{2\sigma_\eta}(\dot a-\pi(a))^2,
\end{equation}
where we dismissed the term $\frac{1}{2}\ln \left\{\sigma_z\sigma_w\sigma_\eta\right\}$ \cite{Kim2018}.
Our specific construct of the IFE encapsulates motor planning explicitly in continuous time via the policy, $\pi(a)$, in the generative models.
Based on the Laplace-encoded IFE, the mathematical statement for the biological FEP is given as
\begin{equation}
\label{FEP}
 \int dt \left\{-\ln p(s)\right\} \le \int dt F(\mu,a;s),
\end{equation}
where the LHS is equivalent to the Shannon uncertainty, $\int ds\left\{-\ln p(s)\right\}p(s)$, under the ergodic assumption, which is assured by the NEQ stationarity of living systems.
The inequality [Eq.~(\ref{FEP})] shows that the upper bound of sensory uncertainty can be estimated by minimizing the time integral of $F$ over a temporal horizon.
Accordingly, if we regard the integrand $F$ as a Lagrangian, the systematic framework of the Hamilton principle can be employed to implement the minimization scheme \cite{Landau1976}.
Next, we cast Eq.~(\ref{Laplace-encoded2}) to a weighted summation of the quadratic terms:
$F = \frac{1}{2}\sum_i m_i\varepsilon_i^2\quad (i=w,z,a)$,
which can be expressed as a total time derivative that does not affect the resulting BM \cite{Landau1976}.
In the summation, we defined the notations $\varepsilon_i$ as
\begin{eqnarray}
\label{pred-errors}
\varepsilon_w &\equiv& \dot\mu-f(\mu),\nonumber\\
\varepsilon_\eta &\equiv& \dot a-\pi(a),\\
\varepsilon_z &\equiv& s-g(\mu,a),\nonumber
\end{eqnarray}
which represent the \textit{prediction errors} involved in state, motor, and sensory inferences, respectively.
Additionally, the weight factors, $m_w,\ m_\eta,\ {\rm and}\ m_z$, are defined through the variances as
\begin{equation}
\label{itmass}
m_w\equiv 1/\sigma_w,\ m_\eta\equiv 1/\sigma_\eta,\ {\rm and}\ m_z\equiv 1/\sigma_z,
\end{equation}
where $m_i$ may be considered as a metaphor for the neural \textit{inertial masses}.
The neural masses correspond to the predictive \textit{precisions} in the standard terminology \cite{BuckleyKim2017}; heavier neural masses lead to more precise predictions.
The IFE F as a Lagrangian, conforming to classical dynamics, can be considered as a function of the \textit{instant trajectories} of $\mu(t)$ and $a(t)$, subject to the time-dependent force, $s=s(t)$.

To exercise the Hamilton principle, we define the \textit{classical Action}, ${\cal S}$, as the time integral of arbitrary trajectories $\mu(t)$ and $a(t)$ in the configurational state space:
\begin{equation}
\label{Action}
{\cal S}[\mu(t),a(t);t)]=\int_{t_0}^t dt^\prime F\left(\mu(t^\prime),a(t^\prime);s(t^\prime)\right),
\end{equation}
where $t_0$ is the initial time, and $\tau\equiv t-t_0$ is the temporal horizon of the relevant biological process.
The initial time can be chosen either in the past, that is, $t_0\rightarrow -\infty$, or at present, that is, $t_0=0$.
In the former, $t$ is the present time, whereas in the latter, $t$ is the future time.
Hence, active inference of the living systems mathematically corresponds to varying ${\cal S}$, subject to the sensory stream, to find an optimal trajectory in the configurational state space spanned by $\mu$ and $a$.

Further, it is advantageous to consider the brain's RD in phase space rather than configurational space; the phase space is spanned by positions and momenta.
This is because the momentum variables are meaningful \textit{prediction errors} in the brain's message passing algorithms; they are defined via the informational Lagrangian, $F$, as
\begin{eqnarray}
p_\mu &\equiv& \frac{\partial F}{\partial \dot \mu} = m_w(\dot\mu-f),\\
p_a &\equiv& \frac{\partial F}{\partial \dot a} = m_\eta(\dot a-\pi),
\end{eqnarray}
where $p_\mu$ and $ p_a$ are the momentum conjugates corresponding to $\mu$ and $a$, respectively.
Equation~(\ref{pred-errors}) reveals that the momenta, $p_\mu$ and $p_a$, are indeed the prediction errors, $\varepsilon_\mu$ and $\varepsilon_\eta$, weighted by the neural masses, $m_w$ and $m_z$, respectively.
The purposive Hamiltonian, $H$, can be obtained by performing the Legendre transformation $H \equiv p_\mu\dot \mu + p_a\dot a - F$.
After straightforward manipulation, we obtain the Hamiltonian function:
\begin{equation}
\label{Hamiltonian}
H(\mu,a,p_\mu,p_a;s) = \frac{1}{2m_w}p_\mu^2 + \frac{1}{2m_\eta}p_a^2 + p_\mu f(\mu) + p_a \pi(a) -\frac{1}{2}m_z\varepsilon_z^2,
\end{equation}
which is a generator of time evolution in neural phase space.
The function H is specified in the cognitive phase space spanned by the four-component column vector, $\Psi$, in the present single-column formulation, whose components are defined as
\[\Psi^T=(\Psi_1,\Psi_2,\Psi_3,\Psi_4)\equiv (\mu,a,p_\mu,p_a),\]
where $\Psi^T$ is the transpose of $\Psi$.
Having determined the Hamiltonian, the Bayesian mechanical equations of motion (termed as BM) can be abstractly written in the symplectic representation as
\begin{equation}
\label{Hameq}
\dot\Psi_i = -J_{ij}\frac{\partial H}{\partial \Psi_j},
\end{equation}
where the block matrix $J$ is defined as
\[
J \equiv
\left(\begin{array}{cc}
0 & -{\bf 1}\\
{\bf 1} & 0
\end{array}\right),
\ {\rm where}\
{\bf 1}=
\left(\begin{array}{cc}
1 & 0\\
0 & 1
\end{array}\right).
\]
Specifically, we unpack Eq.~(\ref{Hameq}) and explicitly display the outcome:
\begin{eqnarray}
\dot\mu &=& \frac{1}{m_w}p_\mu+f(\mu), \label{BM1}\\
\dot a &=& \frac{1}{m_\eta}p_a+\pi(a), \label{BM2}\\
\dot p_\mu &=& -p_\mu\frac{\partial f}{\partial \mu} -m_z(s-g)\frac{\partial g}{\partial \mu},\label{BM3}\\
\dot p_a &=& -p_a\frac{\partial \pi}{\partial a} -m_z(s-g)\frac{\partial g}{\partial a}, \label{BM4}
\end{eqnarray}
which are a coupled set of differential equations that are nonlinear, in general.

The preceding Eqs.~(\ref{BM1})--(\ref{BM4}) comprise the BM of the brain variables, which execute the RD of the Bayesian perception and motor inference in the brain.
The BM was attained by applying the Hamilton principle, for which we adopted the Laplace-encoded IFE as an informational Lagrangian and derived the Hamiltonian to generate the equations of motion.
Our latent variables are the neural representations ($\mu$,$a$) and their conjugate momenta ($p_\mu$,$p_a$); they span the reduced-dimensional neural manifold.
The momenta represent the prediction errors neurophysiologically encoded by the error units in the neuronal population.
{Below, we describe two significant features of the latent dynamics, governed by the BM, subjected to the time-varying sensory input, $s(t)$.
\begin{itemize}
\item[(i)] Equations~(\ref{BM1})--(\ref{BM4}) suggest that the brain mechanistically executes the cognitive operation, which reflects Schr{\"o}dinger's suggestion of an organism as a mechanical work \cite{Schrodinger1944}. Our derived BM addresses the continuous-state implementation of IFE minimization in continuous time, which contrasts common discrete-time approaches \cite{Friston2017,DaCost2020,Sajid2021,Smith2022}. We considered that biological phenomena are naturally continuous and, thus, continuous representations better suit perception and behavior.
\item[(ii)] BM in symplectic form [Eq.~(\ref{Hameq})] represents the gradient-descent (GD) on the Hamiltonian function. However, under nonstationary sensory inputs, the multi-dimensional energy landscape is not static, but incurs time dependence. Accordingly, the presented BM naturally facilitates fast dynamics beyond the quasi-static limit implied by the usual GD methods. In addition, it does not invoke the concept of higher-order motions in the conventional framework \cite{FristonAo2012}; accordingly, our theory is not limited by the issue of \textit{average flows} vs \textit{the rate of change of the average} \cite{Aguilera2021}.
\end{itemize}

\section{Numerical study of BM}
\label{Numerics}
In this section, we numerically develop the latent dynamics of the brain's sensorimotor system resulting from the Hamilton principle-based FE minimization formulation.
For simplicity, we consider a homogeneous, but time-dependent, sensory input, such as nonstationary light intensity or temperature, at the receptors, which emits a motor output innervating the effectors that alter the sensory observation.
There are approximately 150,000 cortical columns in the mammalian neocortex, and each cortical column exhibiting a six-laminae structure may be considered as an independent sensorimotor system \cite{Mountcastle1997,Hawkins2017}.
Our simple model features the double closed-loop microcircuitry delineated in Fig~\ref{Fig1} within a single column, which constitutes the basic computational unit of canonical cortical circuits in an actual large-scale brain network \cite{Bastos2012}.

The generative map, $g$, and functions, $f$ and $\pi$, are unknown; they may be nonlinear or even undescribable within ordinary mathematics.
Here, we exploit the linear models assuming the generic structure:
\begin{eqnarray}
g(\mu,a;\theta_g) &=& \theta_g^{(0)} + \theta_g^{(1)}\mu + \theta_g^{(2)}a, \label{gen1}\\
f(\mu;\theta_f) &=& \theta_f^{(0)} + \theta_f^{(1)}\mu, \label{gen2}\\
\pi(a;\theta_\pi) &=& \theta_\pi^{(0)} + \theta_\pi^{(2)}a, \label{gen3}
\end{eqnarray}
where $\theta_\alpha^{(i)}$ ($\alpha=f,g,\pi$) are the parameters that are to be learned and encoded as long-term plasticity in the neural circuits.
We have included the term $\theta_g^{(2)}a$ in Eq.~(\ref{gen1}), which facilitates the \textit{additive} motor-inference mechanism of the sensory data; additionally, $\theta_g^{(1)}$ and $\theta_g^{(2)}$ magnify or demagnify sensory prediction and motor emission by the internal state, $\mu$, and motor state, $a$, respectively; $\theta_g^{(0)}$ denotes the background error in the measurements.
The constant terms $\theta_f^{(0)}$ and $\theta_\pi^{(0)}$ in Eqs.~(\ref{gen2}) and (\ref{gen3}) specify the prior beliefs on the state and motor expectations, respectively; the coefficients $\theta_f^{(1)}$ and $\theta_\pi^{(2)}$ modulate the relaxation times to the targets.
In addition to these seven parameters $\theta_\alpha^{(i)}$, there appear three neural masses, $m_\alpha$, in the BM unpacked in Eqs.~(\ref{BM1})--(\ref{BM4}).
Hence, the proposed parsimonious BM still encloses 10 parameters, which define a multidimensional parameter space to explore for learning.
The learning problem was not pursued in this study but should be explored in future investigations.
Here, we focus on the active inference problem, assuming that the optimal parameters were already learned or \textit{amortized} over the developmental and evolutionary time scales; these parameters are assumed to be shared for generating present and future sensory data.

By substituting the generative functions given in Eqs.~(\ref{gen1})--(\ref{gen3}) into Eqs.~(\ref{BM1})--(\ref{BM3}), the BM of the state vector, $\Psi$, can be concisely expressed as
\begin{equation}
\label{linearBM}
\dot \Psi + {\cal R}\Psi = {\cal I},
\end{equation}
where the relaxation matrix, $\cal R$, is
\begin{equation}\label{relaxR}
{\cal R} =
\left(
\begin{array}{cccc}
-\theta^{(1)}_{f}            & 0              &   -m_\omega^{-1}          & 0 \\
0           & -\theta^{(2)}_\pi               &0                         & -m_\eta^{-1} \\
-m_z {\theta^{(1)}_g}{\theta^{(1)}_g}         & -m_z \theta^{(1)}_{g}\theta^{(2)}_{g}   & \theta^{(1)}_{f}  & 0 \\
-m_z \theta^{(1)}_g \theta^{(2)}_g   & -m_z {\theta^{(2)}_g}{\theta^{(2)}_g}        & 0          & \theta^{(2)}_{\pi}
\end{array}
\right)
\end{equation}
and the source term, ${\cal I}$, on the right-hand side (RHS) is
\begin{equation}\label{sourceS}
{\cal I}(t) =
\left(\begin{array}{c}
\theta^{(0)}_f \\
\theta^{(0)}_\pi \\
-m_z \theta^{(1)}_gs(t)  + m_z \theta^{(0)}_g \theta^{(1)}_g \\
-m_z \theta_g^{(2)}s(t) + m_z \theta^{(0)}_g \theta_g^{(2)}
\end{array}\right).
\end{equation}
Note that the time-dependence in the source term ${\cal I}$ occurs through the sensory inputs, $s$.
The general solution for Eq.~(\ref{linearBM}) can be formally expressed by direct integration as
\begin{equation}
\label{formalSol}
\Psi(t) = e^{-{\cal R}t}\Psi(0) + \int_0^t dt^\prime e^{-{\cal R}t^\prime}{\cal I}(t-t^\prime).
\end{equation}
The first term on the RHS of Eq.~(\ref{formalSol}) describes the homogeneous solution for an initial condition of $\Psi(0)$, and the second term is the inhomogeneous solution driven by the source, ${\cal I}(t)$, manifesting the history-dependent feature.
The solution represents the brain's cognitive trajectory in action while continuously perceiving the sensory inputs, $s(t)$.

\begin{figure*}[t]
\begin{center}
\includegraphics[width=2.0in]{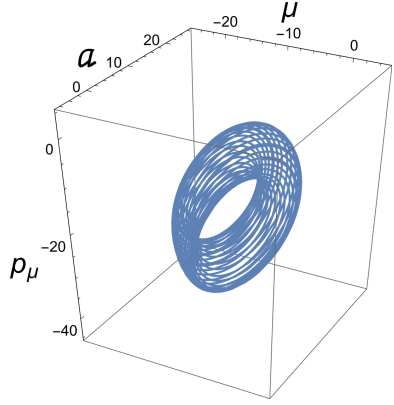}
\caption{Spontaneous attractor: For illustrational purposes, we depict the attractor in the 3D state space spanned by $({\rm Re}[\mu],{\rm Re}[a],{\rm Re}[p_\mu])$ with instantaneous other variables; the attractor center, $\Psi_c$, is positioned at $(-10,10,-20)$. The full attractor evolves in the hyper space spanned by the eight components of complex vector, $\Psi$; in our model, there are the four types of neuronal units $(\mu,a,p_\mu,p_a)$ in a single cortical-column, each of which is allowed to be a complex variable. [Data are in arbitrary units.]}
\label{Fig1}
\end{center}
\end{figure*}
In the long-time limit, $t\rightarrow \infty$, we mathematically predict that the trajectory in the state manifold will fall onto either a fixed point, spiral node or repeller, satisfying $\dot \Psi_{st}=0$ or a limit cycle about a center satisfying $\dot \Psi_{st}=-i\omega\Psi_{st}$, where $\omega$ is an angular frequency characterizing \textit{stationarity}\footnote{In this study, we distinguish between the concept of a stationary state and a steady state: a steady state is the general term indicating the limiting state as $t\rightarrow \infty$, whereas a stationary state is the specific steady state where an oscillatory time dependence remains.}.
The details of the solution's approach to a steady-state will be determined from the eigenvalue spectrum of the matrix ${\cal R}$ and time-varying feature of $s(t)$.
We denote the eigenvalues and eigenvectors by $\lambda(\equiv i\omega)$ and $\phi$, respectively, and set up the eigenvalue problem:
\[
{\cal R}\phi_\alpha=\lambda_\alpha\phi_\alpha.
\]
The trace and determinant are invariant under a similarity transformation; accordingly, the ensuing eigenvalues must satisfy:
\begin{eqnarray}
\sum_\alpha\lambda_\alpha &=& {\rm tr}({\cal R})=0,\\
\prod_\alpha \lambda_\alpha &=& \det({\cal R})\nonumber\\
& =& \theta^{(1)}_{f}\theta^{(1)}_{f}\theta^{(2)}_\pi\theta^{(2)}_\pi + \frac{m_z}{m_w} \theta^{(1)}_{g}\theta^{(1)}_{g}\theta^{(2)}_\pi\theta^{(2)}_\pi + \frac{m_z}{m_\eta}\theta^{(1)}_{f}\theta^{(1)}_{f}\theta^{(2)}_{g}\theta^{(2)}_{g}.
\end{eqnarray}
The eigenvalues form the Lyapunov exponents in the finite-dimensional manifold and characterize the dynamical behavior of the state vector near an attractor.
Because of the multi-dimensionality of the parameter space, it is not ideal to extract the eigenvalue properties analytically from the trace and determinant conditions.
Accordingly, informative constraints on the parameters must be determined on the heuristic basis.
In this study, we numerically searched for parameters that led to pure-imaginary eigenvalues, thereby entailing stationary attractors.

\subsubsection*{Numerical result I: Spontaneous dynamics\\}
We first consider the \textit{spontaneous dynamics} of the brain evolved from the particular solution in Eq.~(\ref{formalSol}) with null sensory inputs in our proposed BM.
The formal representation for the spontaneous trajectory, $\Psi_{sp}(t)$, can be obtained by direct integration as
\begin{equation}
\label{Spontaneous}
\Psi_{sp}(t) = \Psi_c -{\cal R}^{-1}e^{-{\cal R}t}{\cal I}_{sp},
\end{equation}
where the constant vector, $\Psi_c$, is specified as $\Psi_c = {\cal R}^{-1}{\cal I}_{sp}$, where ${\cal I}_{sp}$ is the inhomogeneous term solely from the internal driving sources without the sensory inputs, that is, $s=0$ [see Eq.~(\ref{sourceS})].

In Fig.~\ref{Fig1}, we depict the trajectories generated assuming a set of parameters in the neural generative models [Eqs.~(\ref{gen1})--(\ref{gen3})] as\footnote{These parameter values were selected in an ad hoc manner through numerical inspection to produce a dynamic attractor; therefore, the latent dynamics of the cognitive vector, $\Psi$, is evolved in the extended, complex-valued phase space in the present manifestation.}
\begin{eqnarray*}
    &&(\theta_g^{(0)},\theta_f^{(0)},\theta_\pi^{(0)})= (0,10,10),\\
    &&(\theta_g^{(1)},\theta_f^{(1)})=(2e^{i\pi/2},-1),\\
    &&(\theta_g^{(2)},\theta_\pi^{(2)})=(e^{i\pi/2},e^{i\pi/2}).
\end{eqnarray*}
In addition, the neural inertial masses were assumed to have values of
\[(m_z,m_w,m_\eta)=(1,1,1). \]
The major numerical observations are as follows.
The brain's spontaneous trajectory occupies a limited region in the state space around a \textit{center}, $\Psi_c$, which describes a \textit{dynamic attractor} forming the brain's resting states before sensory influx occurs.
The center is specified by the internal parameters, that is, the generative parameters and neural masses.
We numerically checked that the position of $\Psi_c$ varies with the values of neural masses and the brain's prior belief on the hidden causes of the sensory input and motor state.
We also confirmed that the size of attractors is affected by the generative parameters and neural masses.

\subsubsection*{Numerical result II: Passive recognition dynamics\\}
\begin{figure*}[t]
\begin{center}
\includegraphics[width=0.3\textwidth,height=0.27\textwidth]{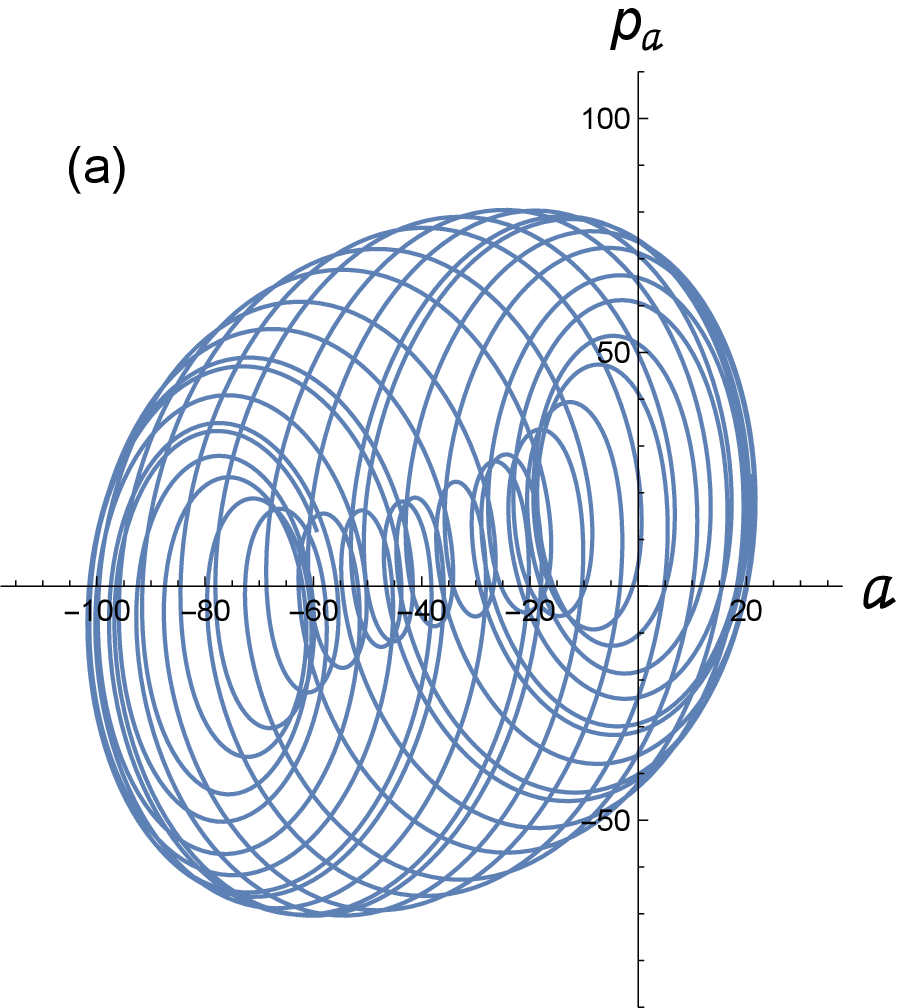}
\includegraphics[width=0.25\textwidth,height=0.27\textwidth]{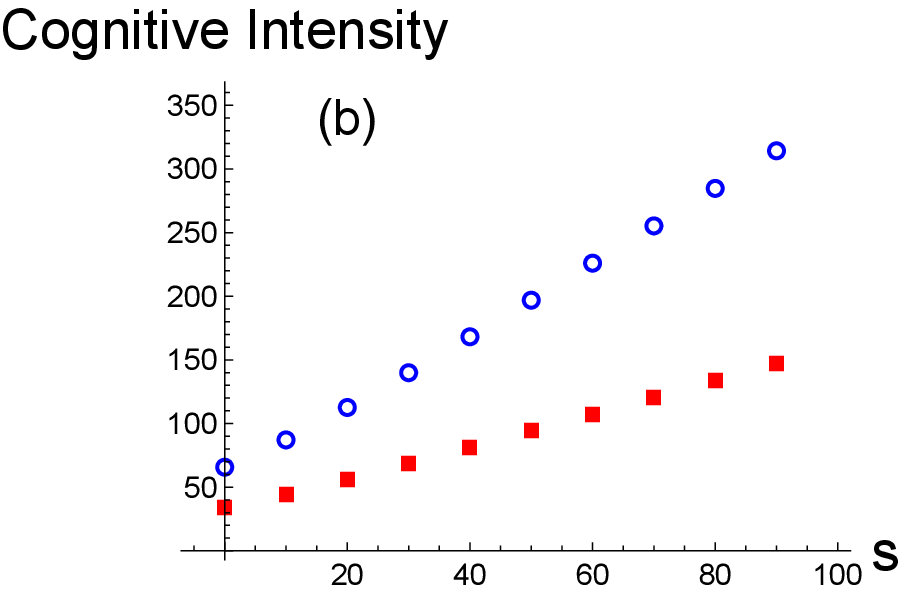}
\caption{Latent dynamics under static sensory inputs: (a) Attractor developed from a resting state, $\Psi(0)$, and driven by the static input $s=100$, using the same parameter values as in Fig.~\ref{Fig1}; the initial state was chosen from the spontaneous states in Fig.~\ref{Fig1}, and for illustrational purposes, the attractor is depicted in the two-dimensional state space spanned by $({\rm Re}[\Psi_2],{\rm Re}[\Psi_4])$. (b) Cognitive intensity, $|\Psi_c|^2$, vs sensory input, $s$. The filled squares are the results from the neural inertial masses $(m_z,m_w,m_\eta)=(10,1,10)$ and open circles are the results from $(m_z,m_w,m_\eta)=(1,1,1)$; the numerical values for the other generative parameters are the same as those used in Fig.~\ref{Fig1}. [Data are in arbitrary units.]}
\label{Fig2}
\end{center}
\end{figure*}
To demonstrate passive perception, we exposed the resting brain to a \textit{static} sensory signal; that is, we inserted $s={\rm constant}$ in Eq.~(\ref{sourceS}).
In this case, the formal solution Eq.~(\ref{formalSol}) can be reduced to
\begin{equation}
\label{Passive}
\Psi(t) = e^{-{\cal R}t}\Psi(0) + \Psi_c -{\cal R}^{-1}e^{-{\cal R}t}{\cal I},
\end{equation}
where, on the RHS, the first term specifies the homogeneous transience of the initial resting state, $\Psi(0)$, second term, $\Psi_c$, denotes the center of attractors, and last term describes the dynamic development from the inhomogeneous source, ${\cal I}(s)$.
In contrast to the spontaneous attractors, the location of center depends on the sensory input, $s$: $\Psi_c = \Psi_c(s) = {\cal R}^{-1}{\cal I}(s)$.

We performed numerical integration and obtained the stationary attractor in the presence of static sensory inputs.
Thus, we confirmed that the attractor behaved similarly as in the spontaneous case, but with a shift of the center because of the nonzero sensory stimulus.
The outcome is presented in Fig.~\ref{Fig2}.
Figure~\ref{Fig2}(a) shows a typical attractor in the two-dimensional state space, which is evolved from a spontaneous state shown in Fig.~\ref{Fig1}.
In addition, in Fig.~\ref{Fig2}(b), we show the change in the \textit{cognitive intensity}, {$|\Psi_c(s)|^2$, with respect to sensory inputs, $s$, which is defined as
\[|\Psi_c(s)|^2 \equiv \Psi_c\Psi_c^*.\]
Given a sensory stimulus, we numerically observe that the cognitive intensity is weaker for a larger inertial mass.
The neural inertial masses represent the inferential precision in the internal models; accordingly, the result shows that less cognitive intensity is required when the internal model is more precise in perceptual inference.
The cognitive intensity may be used as a quantitative measure of awareness or attention in phycology.
Our intensity measure is closely related to neuroimaging analysis \cite{Kuzma2019}, where the neural response to sensory inputs was analyzed as the energy-level change associated with information encoding.

\subsubsection*{Numerical result III: Active recognition dynamics\\}
We considered the nonstationary sensory input, $s(t)$, that renders the time-dependent driving ${\cal I}$ [Eq.~(\ref{sourceS})] in the latent dynamics: the sensory receptors are continuously elicited, and the brain engages in online computation to integrate the BM.
For numerical purposes, we assumed the salient feature of sensory signal, $s(t)$, as a sigmoid temporal dependence:
\begin{equation}
\label{sigmoid}
s(t)=\frac{s_\infty}{1+e^{-k(t-t_m)}},
\end{equation}
where $t_m$ indicates the time when the sensory intensity reaches the midpoint and $k$ adjusts the stiffness of transience in approaching the limiting value, $s(t)\rightarrow s_\infty$.
The sigmoidal sensory inputs are depicted as a function of time in Fig.~\ref{Fig3}(a).

\begin{figure}[t]
\begin{center}
\includegraphics[width=0.3\textwidth,height=0.27\textwidth]{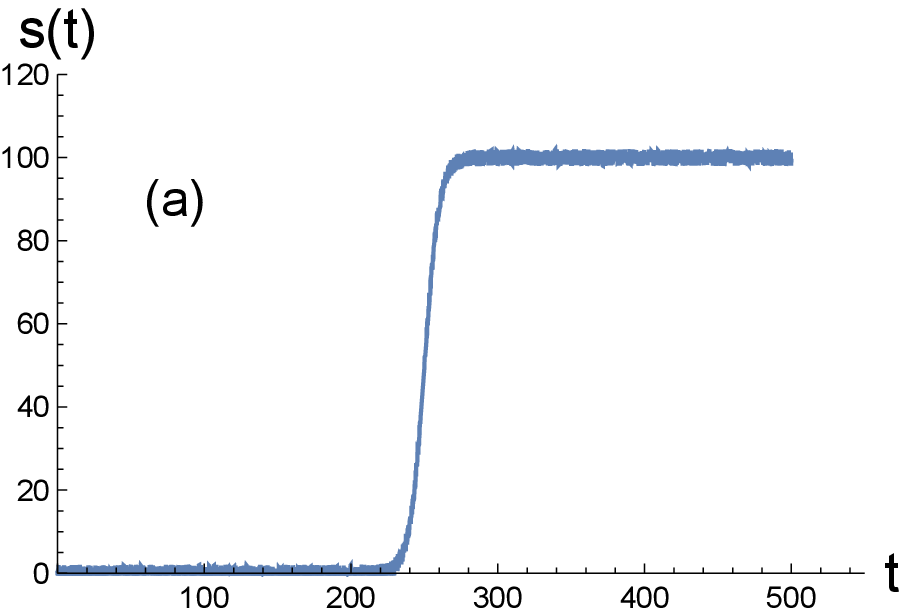}
\includegraphics[width=0.3\textwidth,height=0.27\textwidth]{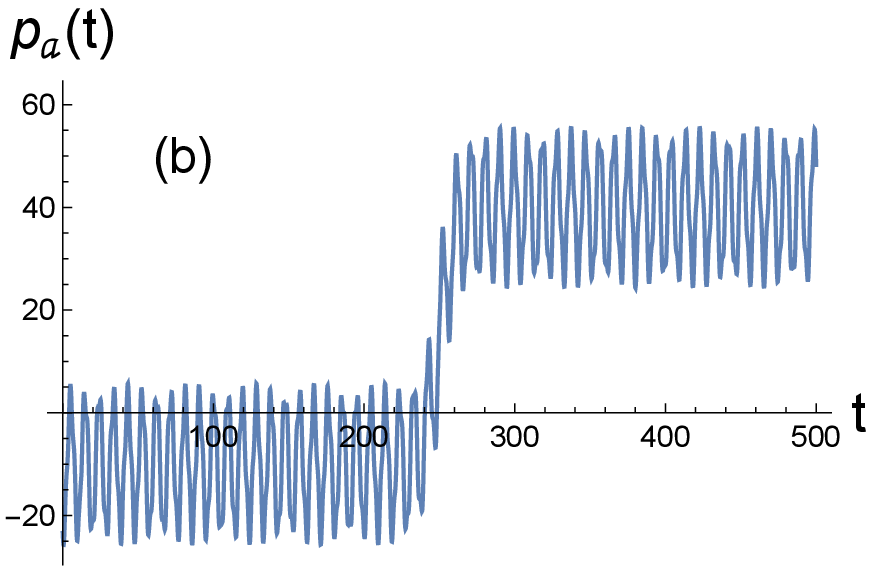}
\caption{Active dynamics under time-dependent sensory inputs: (a) Salient feature of streaming perturbation at the receptor state, $s(t)$; we assume a sigmoid shape for the temporal dependence with the saturated value $s_\infty=100$, stiffness $k=0.2$, and mid-time $t_m=500$. (b) Motor inference of the sensory signals; the BM was integrated using the same parameter values as in Fig.~\ref{Fig2} for the generative parameters and neural masses. [All curves are in arbitrary units.]}
\label{Fig3}
\end{center}
\end{figure}
We numerically integrated Eqs.~(\ref{BM1})--(\ref{BM4}) assuming the same initial state selected for the data shown in Fig.~\ref{Fig2}, subject to the sensory stream presented in Fig.~\ref{Fig2}(a).
In Fig.~\ref{Fig3}(b), we illustrate the imaginary part of the motor state, $a(t)$, in continuous time, which is the online outcome of active inference of the sensory input.
For illustrational purposes, we adopted the sigmoid shape for the temporal dependence with a saturated value of $s_\infty=100$, stiffness of $k=0.2$, and mid-time of $t_m=500$.
The results suggest that the motor state aligns with the sensory variation and successfully infers the sharp change in the sensory input around $t=250$.

In addition, Fig.~\ref{Fig4} presents the attractor dynamics at several time steps exhibiting state transition, \textit{dynamic bifurcation}, from a resting state, $\Psi(0)$, to a cognitive attractor, $\Psi(t)$, over time \cite{Izhikevich2007}.
The numerical computation reveals the initial development of the NEQ attractor with passage of time shown in Fig.~\ref{Fig4}(a) and Fig.~\ref{Fig4}(b), which corresponds to the inferential outcome of the lower part of the sigmoid influx depicted in Fig.~\ref{Fig3}(a).
The intermediate attractor in Fig.~\ref{Fig4}(b) repeats the spontaneous attractor presented in Fig.~\ref{Fig1} because the sensory input is nearly null apart from the negligible fluctuation in the present model.
As time elapses from Fig.~\ref{Fig4}(b) to Fig.~\ref{Fig4}(c), the cognitive state begins to escape from the first attractor and build the second attractor.
Eventually, with passage of time shown in Fig.~\ref{Fig4}(c) and Fig.~\ref{Fig4}(d), the dynamic transition between two attractors completes over a relaxation time period, say, $\tau$.
At time $t> \tau$, the stationary attractor can be described by the expansion
\begin{equation}
\Psi(t)=\bar\Psi_c+\sum_\alpha c_\alpha e^{-i\omega_\alpha t}\phi_\alpha,
\end{equation}
where $i\omega_\alpha\equiv \lambda_\alpha$ and $\phi_\alpha$ are the eigenvalues and corresponding eigenvectors of the relaxation matrix, ${\cal R}$, respectively.
The expansion coefficients, $c_\alpha$, are specified by the initial condition, $\Psi(0)$.
The center of mass of the attractor, $\bar\Psi_c$, is specified by ${\cal R}^{-1}{\cal I}_\infty$, where ${\cal I}_\infty$ is the source vector ${\cal I}$ with the saturated sensory input, $s_\infty$.
The shift of the center between two stationary attractors is shown in Fig.~\ref{Fig4}(d).

The concrete example presented above fully accommodates the active inference of a living agent inferring the sensory signal's salient feature and performing feedback motor-inference in the double closed-loop cognitive architecture [See Appendix].
Although the illustration accounts for a single sensorimotor system, our formulation can also handle multiple modalities of sensory inputs posing multisensory perception problems.
Notably, the time-dependent sensory influx, $s(t)$, makes the linear BM nonconservative, which, from a dynamical-systems perspective, serves as a bifurcation parameter.
Our numerical illustration of the dynamic transition from a resting state to a cognitive attractor is relevant to recent studies of cognitive control of behavior in psychiatry \cite{Cui2020,Parkes2021} and stability of conscious states against external perturbations in patients with brain injury \cite{Deco2019,Perl2021}.
\begin{figure}[t]
\begin{center}
\includegraphics[width=4.0in]{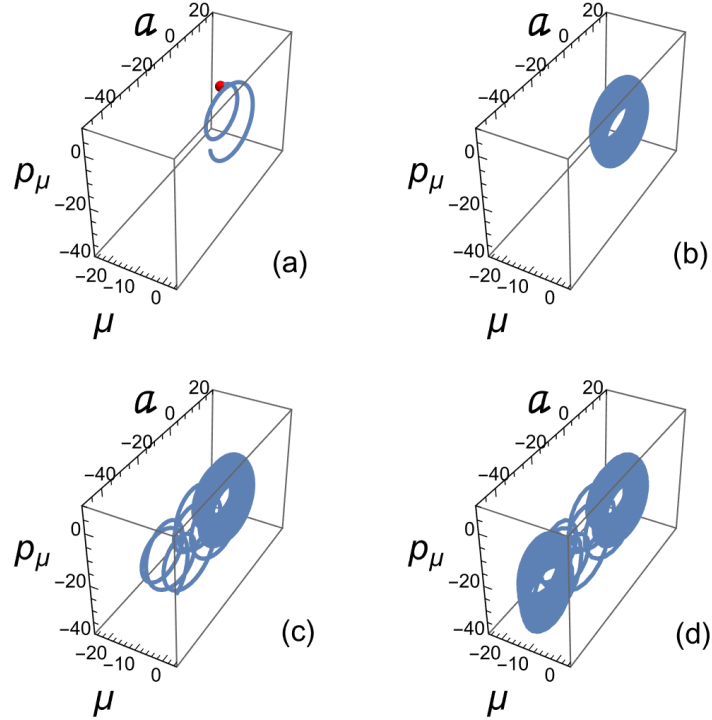}
\caption{Attractor dynamics inferring the nonstationary sensory influx depicted in Fig.~\ref{Fig3}(a): (a) $t=5$, (b) $t=100$, (c) $t=260$, and (d) $t=500$. The trajectory, $\Psi(t)$, results from the direct numerical integration of the BM described by Eqs.~(\ref{BM1})--(\ref{BM4}); the initial state, $\Psi(0)=(-16.9,21.1,-13.3)$, was selected from the spontaneous attractor given in Fig.~\ref{Fig1}. For numerical purposes, the attractor evolution is depicted in the three-dimensional state space spanned by $({\rm Re}[\mu],{\rm Re}[a],{\rm Re}[p_\mu])$. The numerical values adopted for all parameters are the same as those in Fig.~\ref{Fig3}. [Data are in arbitrary units.]}
\label{Fig4}
\end{center}
\end{figure}

\section{Summary and conclusion}
\label{Ending}

This study is based on the consensus that living systems are self-organized into an NEQ stationary state that violates the detailed balance while sustaining physiological and bodily properties.
In a biological context, the thermodynamic second law, the FTs in its modern forms, implies that there is inevitably uncompensated energy in an organism's metabolic processes of maintaining its homeostasis in the environment.
More precisely, the amount of metabolic work is bounded from above by the thermodynamic FE expense.
Efficiency is important in any irreversible phenomena exhibiting the arrow of time, and by extension, in brain work.
We applied modern FTs to a biological agent as an open system and clarified why the concept of the FE is more appropriate than entropy when discussing the question of \textit{What is life?}.
The thermodynamic and neuroscientific FEPs were evaluated based on their respective mathematical inequalities, suggesting the FE bounds as variational objective functions for minimization.
Consequently, we revealed the drawbacks of both principles in accounting for cognitive biological systems and proposed an integrated thermodynamic and Bayesian approach to the biological FEP as a self-organizing principle of life.

The brain states of higher organisms can only be realistically described in terms of probability because of the enormous neuronal degrees of freedom and morphological complexity.
And at the core of the biological FEP are the likelihood and prior densities, making up the G-density, which are thought to be the NEQ probabilities of the physical brain states.
This study argues that the brain dynamics at the mesoscopic, constitutional level are stochastic because of classical negligence, for which time-asymmetric Langevin equations were employed.
The broken time-reversal symmetry was attributed to biological systems being open to the environment.
To statistically describe the brain states, we further used the Markovian approximation in state transitions and adopted the Smoluchowski-Fokker-Planck equation to determine the probability densities of the continuous brain variables.
We viewed the S-F-P equation as a local balance equation for probability and argued that its steady-state solutions furnish the NEQ densities.
The probability flux appearing in the S-F-P equation does not vanish at the brain-environment interface, which reflects that a detailed balance will not be reached in the SS limit, and thus, no standard fluctuation-dissipation theorem is available in the NEQ brain.
Instead, the SS flux resembles the Ampere law in magnetism, resulting from the modified detailed-balance condition and supporting the gradient flow of the NEQ probabilities.

We presented the brain as Schr\"odinger's mechanical machine presiding over predictive regulation of physiology and adaptive behavior of the body.
The BM at the system level is deterministic, indicating that the brain, as a macroscopic physical system, obeys the law of large numbers entailing dimensionality reduction.
In addition, thermal fluctuations from body temperature do not have significant effects on the brain's low-dimensional functions; in other words, the brain is cognitively in its ground state at effective zero temperature.
The IFE was specified in terms of the latent variables that probabilistically encode the environmental and motor states in the brain.
As aforementioned, the encoded probability densities were assumed to be SS solutions to the S-F-P equation or more realistic ones.
Central to our study was the idea that the encoded, online IFE in the brain is a Lagrangian, defining the informational Action.
Based on Hamilton's principle, we found that the brain deterministically conducts allostatic regulation by completing the double closed-loop dynamics of perception and adaptive motor behavior.
We employed a simple model for nonstationary sensory influx and illustrated the development of optimal trajectories in the neural phase space: we numerically observed that the brain undergoes a dynamic transition from a resting state to the stationary attractor, which corresponds to the online inference of the environmental causes in continuous time.
The proposed BM may apply to any generic cognitive processes at the interoceptive, exteroceptive, and proprioceptive levels.

In conclusion, organisms' adaptive sustentation cannot be described within thermodynamic laws and the ensuing TFEP, for which the brain-inspired IFEP provides a promising avenue.
The IFEP, however, utilizes teleological information-theoretic models and then considers the neural bases of those models.
To establish an integrated framework of the organizing principle of life, two rationales of FE minimization and Bayesian inference were hybridized, and the BM directing the brain's latent dynamics of active inference was derived.
Consequently, the brain's perception and motor inference in higher organisms were revealed to operate effectively as Schr{\"o}dinger's mechanical machine.
In addition, we numerically illustrated the attractor dynamics that develops online during a sensory stream in the low-dimensional neural space.

\section*{Acknowledgements}
The author is grateful to J. Kang for providing assistance with the mathematica programming.

\section*{References}
\bibliographystyle{plain}
\bibliography{mybibfile}

\section*{Appendix: Dual structure of perception and motor inference}
Here, we describe a significant feature of our derived BM capturing the dual nature of the sensory and motor inference in the neocortex \cite{Doya2021}, and briefly discuss its relevance to other control theories.

Figure~\ref{Fig5} shows the double-loop architecture of the neural circuitry emerging from the attained BM.
The environmental cause, $\vartheta$, encodes the sensory data, $s$, at the peripheral interface (receptors or input layers), and the brain conducts the variational Bayesian inference that conjointly integrates the double closed-loop dynamics of sensory perception (A) and motor control (B).
Note that the neural units $(\mu,p_\mu,a,p_a)$ are connected by arrows for excitatory driving and by lines guided by filled dots for inhibitory driving.
Loop (A): The state unit, $\mu$, in neuronal population predicts the input, $s$, based on the internal model, $g_1(\mu)$.
The error signal, $\xi_z(\mu)=s-g_1(\mu)$, weighted by the accuracy, $m_z$, of the model, innervates the state-error unit, $p_\mu$, in the population.
The error unit estimates the state by assimilating the discrepancy and sends the feedback signal to the state unit.
Then, the state unit updates its expectation and predicts the sensory input again, which completes the passive perceptual loop.
Loop (B): The motor (effector) unit, $a$, alters the sensory input, $s$, according to the protocol, $g_2(a)$, to promote accurate sensation of the data.
The error signal, $m_z(s-g_2(a))$, acts as a control command to call for an adjustment in the motor-error unit, $p_a$.
Then, the adjusted motor-dynamics transmits the feedback signal to the effector state to further modify the sensory data, which completes the active perceptive loop.
The double closed-loop dynamics concurrently continue until an optimal trajectory, $\Psi(t)$, is fulfilled in the neural hyper-phase space, which corresponds to optimizing the informational classical Action, ${\cal S}$, defined in Eq.~(\ref{Action}).

\begin{figure}[t]
\begin{center}
\includegraphics[width=4.0in]{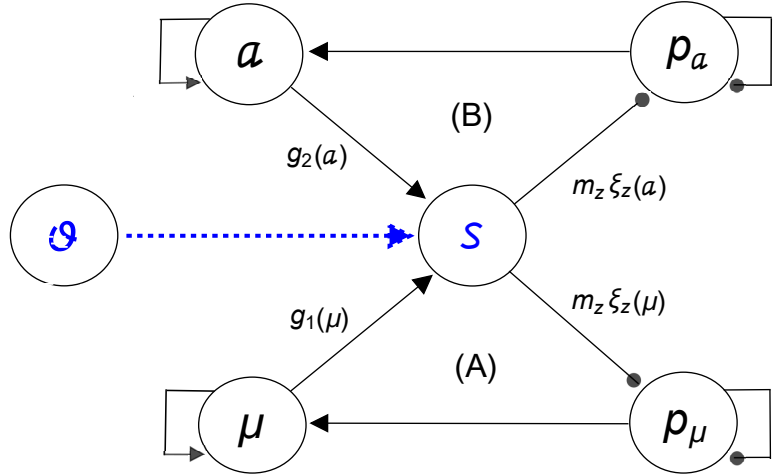}
\caption{Schematic of the neural circuitry exhibiting the double closed-loop architecture, which emerges from the Bayesian mechanics prescribed by Eqs.~(\ref{BM1})--(\ref{BM4}).}
\label{Fig5}
\end{center}
\end{figure}

Our Hamiltonian formulation renders the sensory-driving term, $s-g(\mu,a)$, to appear explicitly in the BM [see Eqs.~(\ref{BM3}) and (\ref{BM4})].
Its role is similar to the unsupervised updating rule in the reinforcement-learning framework \cite{Sutton1998}; specifically, it resembles the continuous control signal in the optimal control theory described by the Hamilton-Jacobi-Bellman equation \cite{Todorov2006}.
The sensory-discrepancy signal not only affects the prediction error, $p_\mu$, in the state prediction [Eq.~(\ref{BM3})], but also the prediction error, $p_a$, of the motor inference  [Eq.~(\ref{BM4})]; this interrelation provides the neural mechanism for adaptive motor feedback via Eq.~(\ref{BM2}).
The momenta in our formulation are termed a \textit{costates} in the deterministic optimal control theory.}

Also, the policy, $\pi$, in Eq.~(\ref{BM2}) accounts for the online motor behavior, which prescribes \textit{motor planning} and can accommodate a \textit{situated decision} \cite{Cos2021}.
In the discrete-state formulations, the policy is defined as a sequence of actions or decisions in discrete time \cite{Sajid2021,Isomura2022}, where the authors incorporate the necessary time-dependence directly in the definition of FE.
On the contrary, our continuous-time theory defines the policy as continuous action planning, which we model as the generative function of motor inference.
The time-dependence of policy generates the history-dependent response of the brain's cognitive state; see Eq.~(\ref{formalSol}), in which the time, $t$, can be either at present or in the future.
When the dynamic perception is coupled to categorical-decision making, the mixed continuous-discrete approaches may shape the active inference problems \cite{Parr2018_DisCon}.}

Finally, we have employed a set of simple and specific generative models [Eqs.~(\ref{gen1})--(\ref{gen3})] for a concrete numerical illustration.
In practice, however, the employed models can be readily generalized.
For instance, one may consider an action-dependent generative function, $f(\mu,a;\theta_f)$, which will make the state dynamics [Eq.~(\ref{state-eq})] subjected to actions.
Further investigations using more realistic models are required to learn the implication and utility of our theory for the dual closed-loop dynamics related to the standard control theories.

\end{document}